%% file: WCCI2016New.tex
\documentclass[letterpaper, 10 pt, conference]{ieeeconf}  

\IEEEoverridecommandlockouts                              

\usepackage{cite}
\usepackage{amsmath}
\usepackage{bm}
\usepackage{xcolor}
\usepackage{makecell}
\usepackage{tabularx}
\usepackage{float}
\usepackage{stfloats}
\usepackage{array}
\usepackage{threeparttable}
\usepackage{graphicx}

\input{Macros}
\usepackage[justification=centering]{caption}

\usepackage[ruled]{algorithm2e}
\usepackage{multirow}

\makeatletter
\newcommand{\removelatexerror}{\let\@latex@error\@gobble}
\makeatother

\title{\LARGE \bf Probabilistic Human Mobility Model in Indoor Environment}

\author{Bo Tang$^{1}$, Chao Jiang$^{2}$, Haibo He$^{1}$, and Yi Guo$^{2}$ 
\thanks{${}^{1}$Bo Tang and Haibo He are with the Department of Electrical, Computer and Biomedical Engineering at the University of Rhode Island,
Kingston, RI, USA. E-mail: {\tt\small \{btang,he\}@ele.uri.edu}}
\thanks{${}^{2}$Chao Jiang and Yi Guo are with the Department of Electrical and Computer Engineering at the Stevens Institute of Technology,
Hoboken, NJ, USA. E-mail: {\tt\small \{cjiang6,yguo1\}@stevens.edu}.
}
}

\begin{document}

\maketitle
\thispagestyle{empty}
\pagestyle{empty}

\begin{abstract}
Understanding human mobility is important for the development of intelligent mobile service robots as it can provide prior knowledge and predictions of human distribution for robot-assisted activities. In this paper, we propose a probabilistic method to model human motion behaviors which is determined by both internal and external factors in an indoor environment. While the internal factors are represented by the individual preferences, aims and interests, the external factors are indicated by the stimulation of the environment. We model the randomness of human macro-level movement, e.g., the probability of visiting a specific place and staying time, under the Bayesian framework, considering the influence of both internal and external variables. We use two case studies in a shopping mall and in a college student dorm building to show the effectiveness of our proposed probabilistic human mobility model. Real surveillance camera data are used to validate the proposed model together with survey data in the case study of student dorm.
\end{abstract}

\section{Introduction}
Understanding human mobility in indoor environments is important for many applications such as human-robot interaction and cooperation, emergency evacuation, and crowd disaster prevention. For the development of mobile service robots, human mobility modeling can provide dynamic distributions of human indoor activities, and help robots to plan their paths, motion, interaction, and engagement with the humans \cite{tipaldi2011want,Nikhil12,macharet13,tang2016robot}. For example, multiple mobile service robots can be distributed in a reasonable way according to the crowdness of a place in a shopping mall which usually dynamically changes depending on many factors, e.g., sales promotions. Besides, human mobility modeling can provide prior knowledge and predictions of human distribution for robot-assisted emergency evacuation in indoor environments. For example, the dining hall is usually more crowded at lunchtime than other places in a shopping mall, and an emergency guidance robot goes around the crowded place in advance could improve evacuation efficiency when an emergency takes place. 


Modeling human mobility in indoor environments is a challenging task.  In the existing literature, there are two types of methods on human
mobility modeling: individual human trajectory modeling and statistical characteristic modeling. While the former method focuses on modeling individual motion behavior and predicting his/her future possible trajectories in an indoor environment \cite{bennewitz2002learning} \cite{ziebart2009planning}, the latter one aims at investigating statistical characteristics of human trajectories in an outdoor environment, such as
visiting place distributions and staying time distributions. The latter human mobility model outdoor is well studied through empirical data analysis by monitoring human daily activities using GPS data and mobile phone records, where human trajectories were described as random walks, such as Levy Flights or continuous time random walks \cite{brockmann2006scaling, gonzalez2008understanding, song2010modelling, rhee2011levy}.
However, existing human mobility modeling methods using the random walk framework may fail to model human mobility in an indoor environment, due to its space constraints and limited interests of visiting places.

Despite of uncertainties in human mobility modeling, it is not random. Considering places of interest in an indoor environment, e.g., shops in a shopping mall, more customers would be around those places or areas that are of higher interest for customers at a specific time, such as the dining hall during the meal time or the outlet stores with attractive discounts. The list of shops to visit for a customer may also be determined by his/her own
preference and entrance he/she enters. Inspired by the process of human decision making \cite{helbing1995social}, we use both internal and external factors to model human's spatio-temporal behavior. The internal factors are represented by individual characteristics (including age, education level, cultural, etc.), aims, interests and motivations, while the external factors are indicated by environment stimulus and group behaviors. The influence of these internal and external factors (variables) on place selection could be reflected by the staying time of each place, based on the
fact that people have much more interest if he/she stay at a place long time, and less if they don't. 

In this paper, we propose a probabilistic human indoor mobility model considering both internal and external variables to capture the
uncertainty of human intention. Given the internal and external variables of each person, we can calculate the posterior probability of visiting a specific place and the staying time. The human dynamic behavior is then determined by the proposed human indoor mobility model, which can provide dynamic distributions of humans and their staying time distribution at any given time. In our simulations, we conduct two case studies: one is the simulation of human movement in an indoor shopping mall using the proposed mobility model, and the other one is the simulation of human distribution in a student dorm building. In the latter case study, we use the survey data as prior knowledge and apply real-life camera data to validate the performance.

The rest of the paper is organized as follows. In section II, we provide related work in human mobility modeling. In section III, we present our proposed human mobility model for indoor environments. In Section IV, extensive simulations and empirical studies are performed to validate the feasibility of our proposed human mobility model. Finally, conclusions are given in Section V.

\section{Related Work}
A lot of research efforts have been made on understanding human motion behavior for the last few decades. Due to the complex psychological and physical decision-making process, it is known that individual human motion behavior in an unconstrained situation is hard to be understood. Human mobility modeling can be grouped into two categories: \textit{indoor} and \textit{outdoor} human mobility. For outdoor human mobility, recent studies have shown that statistical properties of human mobility can be found from real human motion traces, such as distributions of moving speed and pause. The real human motion traces are usually long-term and long-distance observations obtained by using GPS, cellphones, and WiFi localization technologies \cite{brockmann2006scaling}\cite{gonzalez2008understanding}\cite{rhee2011levy}. It is shown that the outdoor human motion behavior has a truncated power-law distribution. Many random mobility models have also been proposed in terms of the statistical feature of human walking behavior, such as random walk \cite{shlesinger1982random}, Levy flight \cite{brockmann2006scaling}, truncated Levy flight \cite{gonzalez2008understanding}\cite{rhee2011levy}, Pragma
\cite{borrel2005preferential}, hotspots \cite{kim2006extracting}, to name a few.

For human indoor mobility model, typical work includes crowd analysis and evacuation behaviors in panic situations \cite{helbing1995social} \cite{zhan2008crowd}\cite{helbing2000simulating}, and learning and inference of human motion patterns \cite{bennewitz2002learning}\cite{ziebart2009planning}. Crowd analysis and evacuation behaviors focus on the understanding of individual human behaviors in panic crowds. Helbing contributed a series of works on both crowd analysis and evacuation behavior modeling. He presented a formation of pedestrian flows mathematically with a Boltzmann-like model \cite{helbing1993stochastic}, a social force model for pedestrian dynamics in \cite{helbing1995social}, and dynamic behavior simulations in emergency situations \cite{helbing2000simulating}. Some realistic self-organization collective behaviors are observed in the simulation of using social force model. Instead of mathematically modeling the crowd motion, fuzzy methods using agent-based models can be employed to describe human dynamic behaviors. A nice survey on crowd analysis can be found in \cite{zhan2008crowd}. In \cite{pan2006human}, the authors employ
agents to describe non-adaptive crowd behaviors using three levels, including the individual, the interactions among individuals, and the
interaction between the group and environment. Learning and inference of human motion patterns aim to capture the uncertainty of human intention based on Bayesian framework using modern machine learning methods, such as expectation maximization algorithm \cite{bennewitz2002learning}, Bayesian network \cite{tang2015turn}, Markov decision process \cite{ziebart2009planning}, tracking filter \cite{schulz2001tracking}, among others. Learning human motion patterns from previous trajectories and poses can also predict their next trajectories and poses, which is critical for robot-human interaction and cooperation.

While outdoor human mobility can be modeled at a large scale, it is challenging to model human indoor mobility in a much smaller scale for indoor
environments. In this paper, we present our new probabilistic method for human indoor mobility modeling.

\section{Human Indoor Mobility Modeling}
\begin{figure}[!ht]
\centering
\includegraphics[width=0.4\textwidth,natwidth=610,natheight=642]{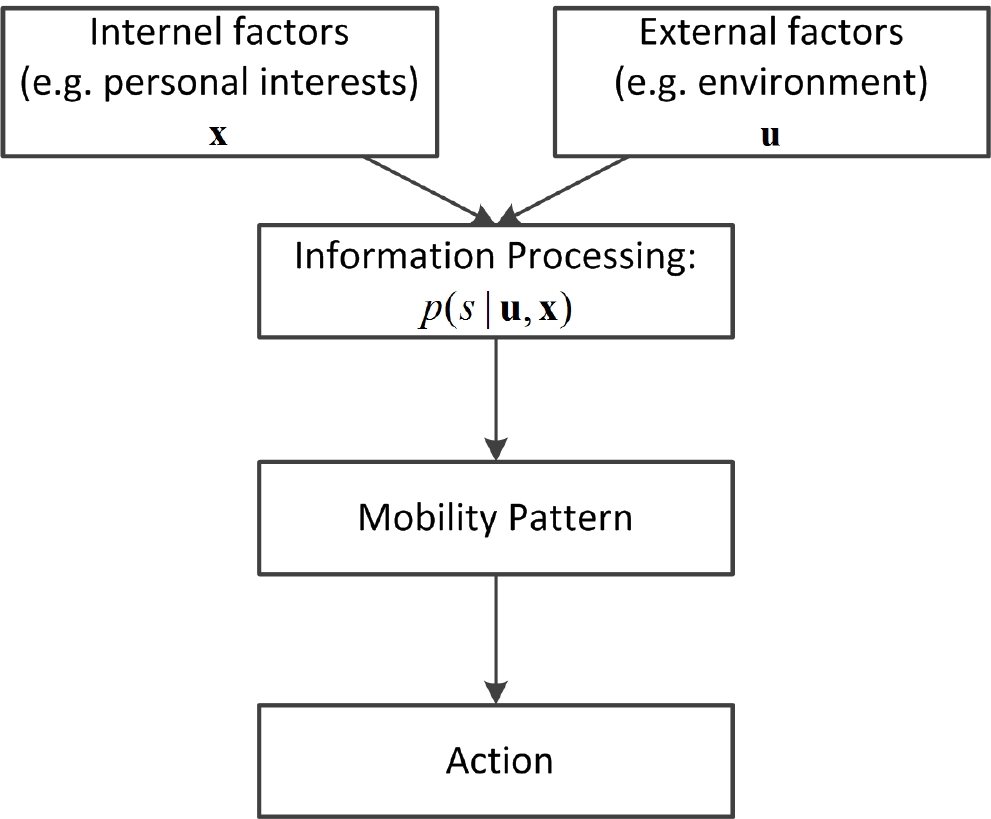}
\caption{The scheme of the proposed probabilistic human mobility model}
\label{process}
\end{figure}

Dynamic human distribution in a constrained environment is usually not deterministic, but stochastic. Considering an indoor environment with places of interest, such as a shopping mall with various kinds of shops or a college student dorm building with a lot of facility rooms, we are interested in knowing the probability of visiting places when one person passes by a specific place. Motivated by the \textit{``latent"} variables and the \textit{``stimulus''} in behavior change process used in the social force model \cite{helbing1995social}, human dynamic behavior in such an indoor environment can be determined by both {\em internal} and {\em external} factors. The internal factors are referred to internal motivations, which include factors driving a person's preference in certain actions of movement, such as their age, gender, income and education level, among others. The external factors are defined as output stimuli, which include external environment conditions  impacting a person's mobility pattern. Given an example of a shopping mall, each customer in the mall may have his/her own habits depending on internal factors, such that he/she could be attracted by a specific shop. Meanwhile, if there is a sale in a certain shop, we can also expect that the customer will move towards that shop. Indeed, our definitions of internal and external factors are well aligned with the hypothesis of the social force model for pedestrian dynamics \cite {helbing1995social}, in which one can assume that a behavioral reaction is caused by a sensory of stimulus that depends on the personal aims and the environment with the objective of utility maximization. The proposed indoor human dynamic mobility model is shown in Fig. \ref{process}.

Assuming there are total $M$ places and $N$ people in an indoor environment, we denote $s_i$ and $\mathbf{u}_i$ as the $i$-th place of interest and its external factors, and $\mathbf{x}_j$ as the $j$-th person with a specific internal state. For the $j$-th person $\mathbf{x}_j$, we define the probability of visiting the $i$-th place as $p(s_i = 1| \mathbf{x}_j, \mathbf{u}_i)$ given the external factors $\mathbf{u}_i$, where $s_i = \{0, 1\}$. $s_i = 1$ indicates that the person visits the $i$-th place, while $s_i = 0$ means not. For the simplicity, we replace $s_i = 1$ with the term $s_i$ in our remaining paper. We have the following assumptions for our human mobility model in normal case: first, we assume that internal and external variables are independent; second, we assume that the prior internal and external distributions $p(\mathbf{x} | s)$ and $p( \mathbf{u} | s )$ for a specific place can be obtained from statistical analysis of historical data.

As the probability of $p(s_i | \mathbf{x}_j, \mathbf{u}_i)$ may vary for different people and different external events or factors, according to the Bayesian theorem, the probability of visiting event can be further written as:
\begin{align}
p(s_i | \mathbf{x}_j, \mathbf{u}_i) = \frac{p(\mathbf{x}_j | s_i, \mathbf{u}_i) p(s_i |
\mathbf{u}_i)}{p(\mathbf{x}_j | \mathbf{u}_i)}
\end{align}

Due to the independence of internal and external
states, we have $p(\mathbf{x}_j | \mathbf{u}_i) = p(\mathbf{x}_j)$ and $p(\mathbf{x}_j | s_i, \mathbf{u}_i) = p(\mathbf{x}_j | s_i)$. Hence, given the external factors for a particular place, we have
\begin{align}
p(s_i | \mathbf{x}_j, \mathbf{u}_i)  & = \frac{p(\mathbf{x}_j | s_i) p(s_i |
\mathbf{u}_i)}{p(\mathbf{x}_j)} \nonumber \\
& = \frac{p(\mathbf{x}_j | s_i) p( \mathbf{u}_i |s_i) p(s_i) }{p(\mathbf{u}_i)
p(\mathbf{x}_j)}
\label{normal_likelihood}
\end{align}
where $p(\mathbf{x}_j | s_i)$ and $p(\mathbf{u}_i |
s_i )$ are the prior probability of internal and external variables for the
$i$-th place, respectively. For the $j$-th people with internal
states $\mathbf{x}_j$ who passes a particular place $s_i$, given the prior
distribution $p(\mathbf{x}_j | s_i)$ and $p( \mathbf{u}_i | s_i )$, we can
calculate the likelihood of event whether he/she visits the place ($s_i = 1$) or
just passes it ($s_i = 0$). Note that the above model can be extended to
the case in which external variable distribution changes over time, i.e.,
$p(\mathbf{u}_i | s_i) = p(\mathbf{u}_i(t) | s_i)$, as we demonstrate in our
simulations. Hence, the above probabilistic model can be used to describe human
dynamic distribution over time.

We model the human spatio-temporal distribution in terms of staying time for the
place people have visited. The staying time in a place for a person can reflect his/her
probability of visiting a place, i.e., the likelihood someone visits a place
given both of his/her internal and external factors. We determine the occurrence of visiting event if the probability exceeds a specific threshold $\tau$, that is to
say the person moves into that particular place. Meanwhile, his/her staying time
$t_i$ is also determined by this probability. We further have
\begin{align}
t_i = \frac{p(s_i | \mathbf{x}_j, \mathbf{u}_i) - \tau}{1 - \tau} \times
t_{\text{max}}
\label{normal_staytime}
\end{align}
where $t_{\text{max}}$ is the maximum staying time in $i$-th place over all
people. In Eq. (\ref{normal_staytime}), we know that if $p(s_i | \mathbf{x}_j, \mathbf{u}_i) =
\tau$, then the staying time is zero, i.e., the person will not visit
this place, and if $p(s_i | \mathbf{x}_j, \mathbf{u}_i) = 1$, then his/her
staying time is $t_{max}$. 

The algorithm of human mobility simulation is
presented in Algorithm \ref{algorithm: human dynamic model}.  In Algorithm
\ref{algorithm: human dynamic model}, the order of all possible visiting places
is initialized as $\mathbf{x}_j.path$. There are three different statuses:
``stay the same place'', ``leave current place'', and ``go to next place''. When
people keep staying on one place with status of ``stay the same place'', we
model their behaviors as a random walking model. Once the staying time is
counted to the expected staying time, the status becomes ``leave current place''
and the destination is updated from $\mathbf{x}_j.path$ for the $j$-th people.
The likelihood is calculated to determine the staying time. When the $j$-th
people go the next place of interest, we employ the social force model \cite{helbing1995social}
to model the individual trajectory. The social force model describes individual
behavior in a crowd using social rules. The trajectory is usually determined by
the self-motivation force and interactive forces with others and walls. While
the self-motivation force determines the desired walking direction and speed, the
interaction forces indicate the avoidance of physical contact with others and
wall. For the details of social force model, the interested reader is referred
to \cite{helbing1995social} for further information. Note that the social force
model deals with human-human interaction while our proposed model focuses on a
higher-level of human mobility such as visiting places and staying time.
Combining both models, we can simulate individuals' trajectories in indoor
environments as shown in the next section. 

\begin{figure}[!t]
 \removelatexerror
  \begin{algorithm}[H]
   \label{algorithm: human dynamic model}
   \caption{Probabilistic Human Mobility Model}
   \textbf{Require}: prior internal and external variable distributions
   $p(\mathbf{x} | s)$ and $p(\mathbf{u}|s)$; \\
   \textbf{Initialize}: \\
   $N$ persons with internal states $X = \left\{\mathbf{x}_{1}, ...
   ,\mathbf{x}_{N} \right \}$, \\
   $M$ places of interests $S = \left \{ s_{1},
   ...
   , s_{M} \right \} $ and corresponding external factors $U =
   \left \{ \mathbf{u}_{1}, \ldots , \mathbf{u}_{M} \right \} $\;
   \For( \textit{Simulation loop} ){$t := 1 \to T$}{
	   \For(){$j := 1 \to N$}{
	      	\If{$\mathbf{x}_j.status$ == ``leave current place''}{
	      		next place $i \leftarrow \mathbf{x}_j.path.next$\;
	      		calculate the likelihood $p(s_i | \mathbf{x}_j, \mathbf{u}_i) $ in Eq. (\ref{normal_likelihood})\;
	      		calculate the staying time $\mathbf{x}_j.stayingTime$ in Eq. (\ref{normal_staytime})\;
	      		$\mathbf{x}_j.status$ $\leftarrow$ ``go to next place''\;}
	      	\If{$\mathbf{x}_j.status$ == ``go to next
	      	place''}{ update the position\;
	      		\If{$\mathbf{x}_j.pos \in Area_i$}{
	      			$\mathbf{x}_j.status$ $\leftarrow$ ``stay the same place''\;
	      			$\mathbf{x}_j.count = 0$\;
	      		}
	      	}
	      	\If{$\mathbf{x}_j.status$ == ``stay the same place''}{
	      		randomly walk\;
	      		$\mathbf{x}_j.count ++ $ \;
	      		\If{$\mathbf{x}_j.count >= \mathbf{x}_j.stayingTime$} {
	      			$\mathbf{x}_j.status$ $\leftarrow$ ``leave current place''\;
	      		}
	      	}	
	      }
   }
   \end{algorithm}
\end{figure}

\begin{figure*}[bp]
\begin{eqnarray}
\label{simulation_prior}
p(\mathbf{x}| s = s_i) & = p(x_{age},x_{income} | x_{edu} = \text{``HS"}, s =
s_i) p(x_{edu} = \text{``HS"}| s = s_i) + \nonumber \\
& \quad p(x_{age},x_{income} | x_{edu} = \text{``BA"}, s = s_i) p(x_{edu} = \text{``BA"} | s = s_i) + \nonumber \\
&  \quad  p(x_{age},x_{income} | x_{edu} = \text{``MA"}, s = s_i) p( x_{edu} = \text{``MA"} | s = s_i) + \nonumber \\
&  \quad p(x_{age},x_{income} | x_{edu} = \text{``PHD"}, s = s_i) p(x_{edu} =
\text{``PHD"} | s = s_i) + \nonumber \\
&  \quad  p(x_{age},x_{income} | x_{edu} = \text{``NONE"}, s = s_i) p(x_{edu} = \text{``NONE"} | s = s_i)
\end{eqnarray}
\end{figure*}

\section{Case Studies and Analysis}

\subsection{Case Study 1: Dynamic Human Distribution in a Shopping Mall}

\begin{figure}[!ht]
\centering
\includegraphics[width=0.3\textwidth,natwidth=610,natheight=642]{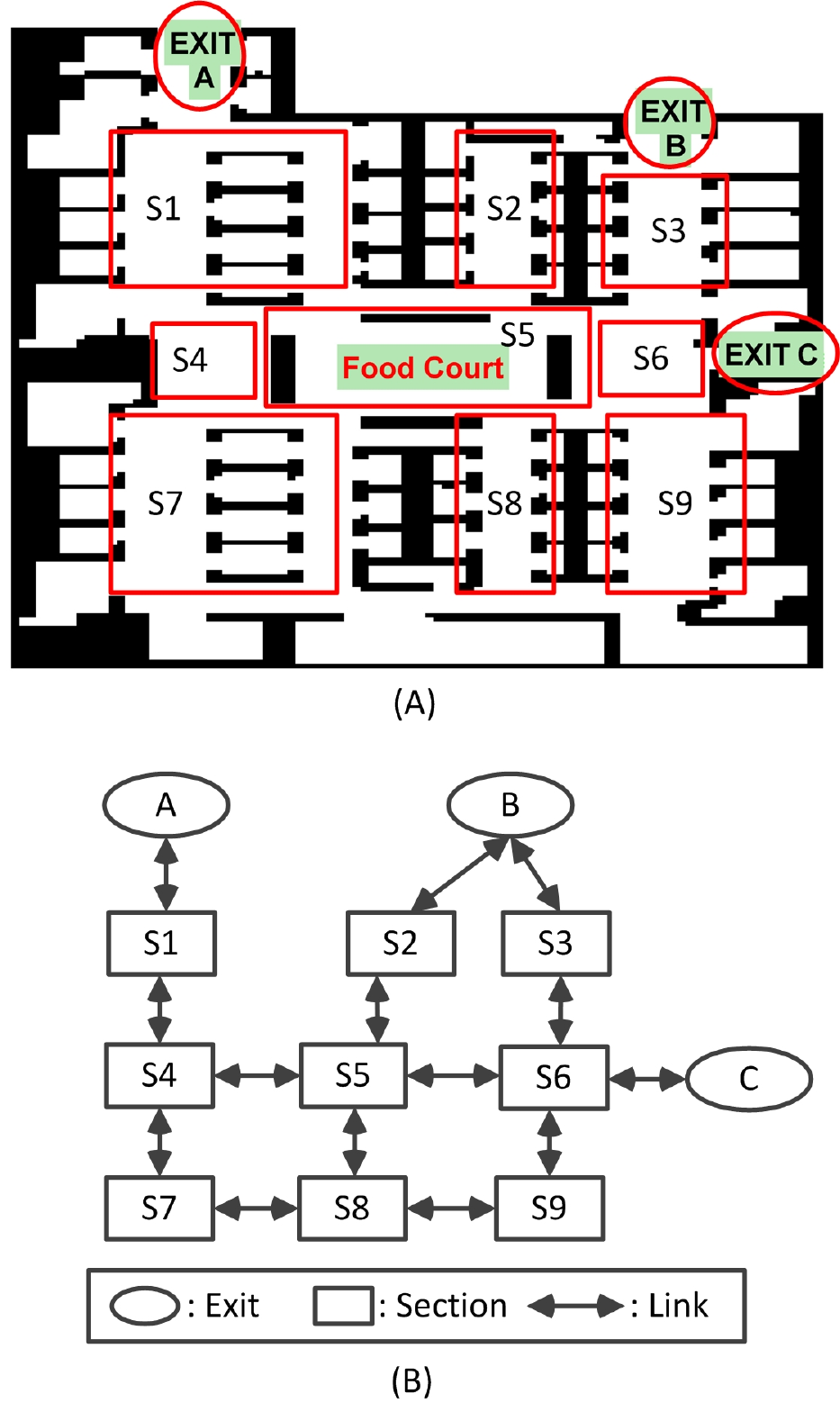}
\caption{(A) The layout of the shopping mall; (B) The network graph $G$ which is a collection of vertices and edges to represent the environment layout.}
\label{map_layout}
\end{figure}

Using the proposed probabilistic human mobility model, we firstly simulate the dynamic distribution in a shopping mall which has $N_e = 3$
entrances/exits and $M = 67$ shops. The layout of this indoor environment is
shown in Fig. \ref{map_layout}(A). The complexity of such indoor environment can
be simplified by a network graph $G$, as shown in Fig. \ref{map_layout}(B), in which the
shopping mall layout is represented as $9$ sections and $3$ exits. The
vertex of network graph $G$ is the centroid of a section, and the edge
connecting two adjacent sections is the distance of these two centroids.

Given the example of shopping mall with the layout shown in Fig.
\ref{map_layout}, we simulate human dynamic motion to illustrate the process
of our human mobility model in an indoor environment with various places of
interest. To simplify the simulation, we consider the age, income, and education
level as internal variables, denoted as $\mathbf{x} = [x_{age}, x_{income},
x_{edu}]$. Notice that more internal variables could be easily added into our
human mobility model. While both $x_{age}$ and $x_{income}$ are continuous
variables, $x_{edu}$ is a discrete variable. The prior joint distribution of
internal variable for the $i$-th shop $p(\mathbf{x}| s = s_i)$ can be written as
Eq. (\ref{simulation_prior}), if five education levels are considered, where
``NONE", ``HS", ``BA", ``MA" and ``PHD" which mean the degree of high school, bachelor, master,
and doctor of philosophy, respectively, and ``NONE" means the person doesn't
graduate from high school. For each shop, we assume this customer's internal
variable distribution is known as prior knowledge that can be estimated
from historical customer data. In our simulation, we use an uniform
distribution to model the distribution of education level $p(x_{edu} | s_i)$ and
Gaussian mixture distributions to model the conditional distributions of
$x_{age}$ and $x_{income}$ given an education level $x_{edu}$, i.e.
$p(x_{age},x_{income} | x_{edu},s_i) \propto \sum_{k=1}^{K_i} w_{i,k}
\mathcal{N}(\bfmu_i, \bfSig_i)$.

\begin{figure*}[tp]
\centering
\includegraphics[width=1\textwidth,natwidth=610,natheight=642]{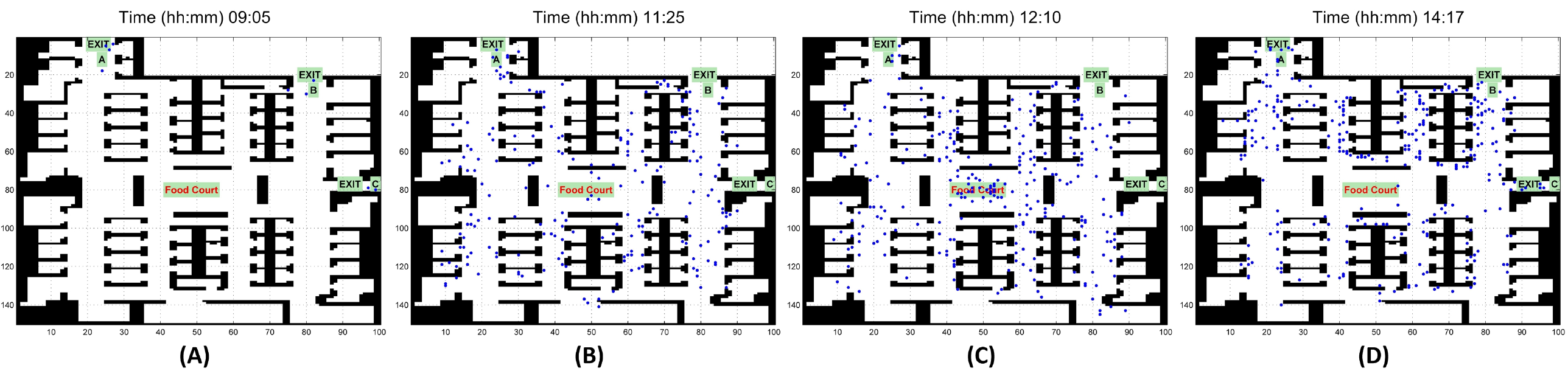}
\caption{The simulation of customer distribution over time for a typical day in
a shopping mall.}
\label{normal_mall}
\end{figure*}

\begin{figure}[!ht]
\centering
\includegraphics[width=0.4\textwidth,natwidth=610,natheight=642]{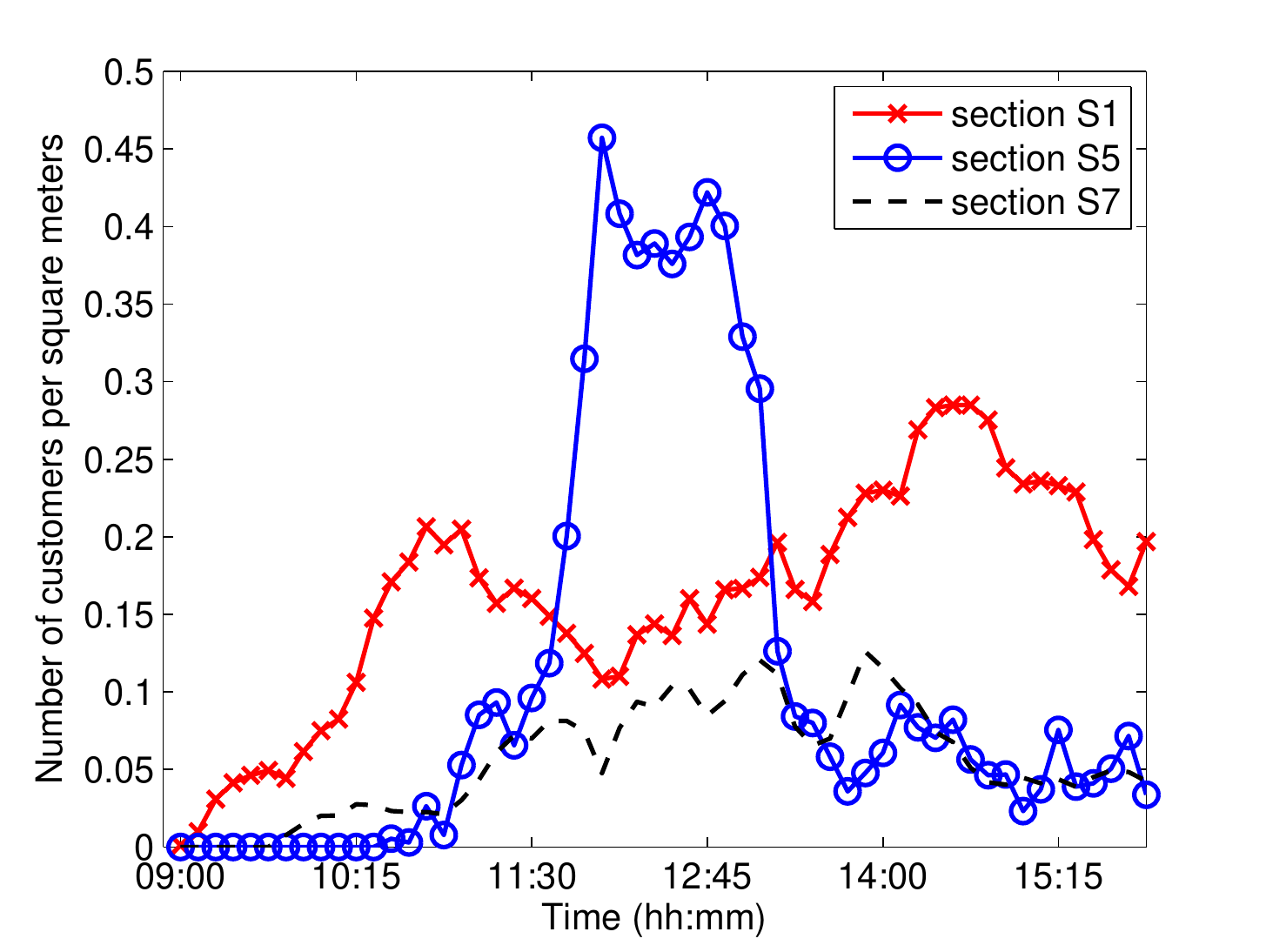}
\caption{Number of people per square meter at different sections in a shopping mall.}
\label{section_density}
\end{figure}

While the internal variables are constant for each specific people, the external
variables may be dynamically changed over time for each shop, such as
limited-time promotional discounts. Another example is the Food Court
in the shopping mall, where we can expect a higher customer distribution
during the lunch and dinner time. In our simulation, we consider dynamic
external variables. For example, there is a peak in the distribution of $p(\mathbf{u} | s)$ in Eq. (\ref{normal_likelihood}) from $11:30 \
AM$ to $1:30 \ PM$ at the Food Court. Using such external variable
distribution, we show the simulation results of customer distribution over time
in Fig. \ref{normal_mall} for a typical day in the shopping mall. In the
simulation, we generate a customer per $15$ seconds with a prior distribution $p(\mathbf{x})$ and uniformly choose an entrance with a probability
$1/N_e$. When the customer finishes shopping, he/she leaves the mall from the
same entrance. To determine whether a customer will enter a specific shop, we
calculate the probability $p(s_i | \mathbf{x}_j, \mathbf{u}_i)$ and compare it
with a threshold $\tau$. If he/she
visits the shop, the staying time is also determined by  $p(s_i | \mathbf{x}_j,
\mathbf{u}_i)$ in Eq. \ref{normal_staytime}. As there is a peak distribution
from $11:30 \ AM$ to $1:30 \ PM$ at the Food Court, we can expect
that more customers are attracted to visit it during this
period. Fig. \ref{section_density} shows the customer density, defined as the
total number of customers per square meter, over time for different sections in shopping mall. One can see several distinct time
series patterns of customer distribution at different sections. For example,
the customer density in section S5, as we expected, increases at lunch time as
many customers visit the Food Court for lunch. Except for the lunch time, there
is always a high customer density in section S1 as it is close to one of
entrances, and a low customer density in section S7 as it is a corner which is
far from all entrances. The patterns of human dynamic
distribution observed from our simulations may find wide applications such as business
reallocation to attract more visitors, crowd flow design to
avoid the occurrence of crowd disaster, robot path planning to guide robots to serve more customers.

\subsection{Case Study 2: Average Facility Usage in a Student Dorm Building}
In this case study, we simulate human dynamic motion in a college dorm building and examine the average usage of different facilities. We consider a real-world college student dorm building, Jonas hall in Stevens Institute of Technology, as an indoor environment in our simulations. The places of interest are various places of this dorm building which include student dorm rooms, halls, RA offices, kitchens, washers/dryers, vending machines, public bath, elevator, front door, back door, and stairs. 

\subsubsection{Prior Knowledge Acquisition Through Survey}
\begin{figure}
\centering
\includegraphics[width=0.4\textwidth,natwidth=610,natheight=642]{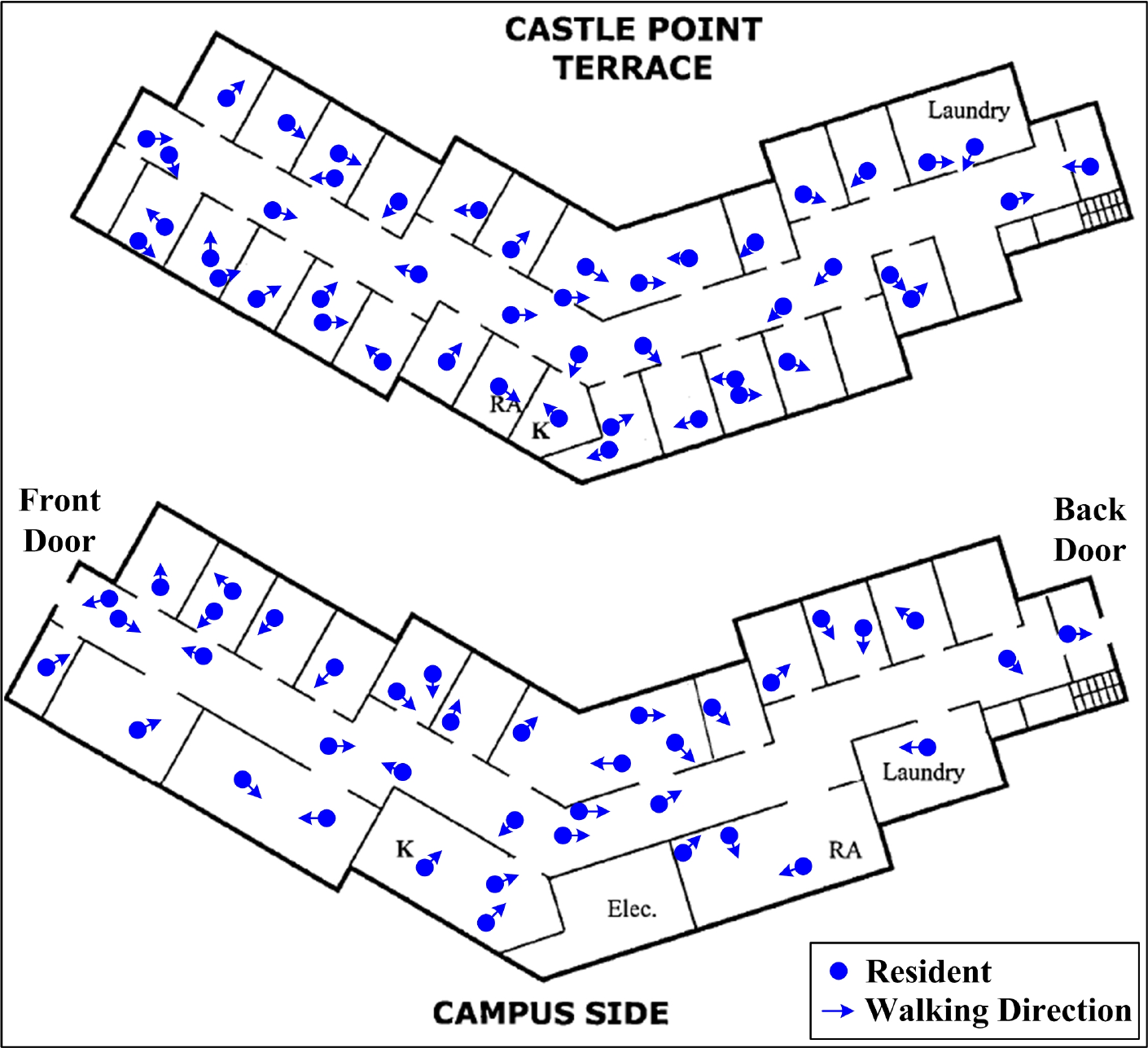}
\caption{Resident distribution in the Jonas Hall building during a typical
weekday on Stevens campus}
\label{campus}
\end{figure}

To acquire the prior knowledge, a survey was created to obtain the statistical information about residents' habits involving time spent in various locations of the dorm building during different times of the day and week \cite{suvery}. There are three floors in this hall, and we only consider the first and the third floor in our survey and simulations, as shown in Fig. \ref{campus}. There are $230$ residents in this dorm building, and $50$ of them are surveyed from these two floors. Three types of days are considered as external factors: weekday, weeknight, and a typical day of weekend. In the survey, we ask students to estimate the average staying time (in minutes) they spend at these $11$ places in a typical weekday, weeknight and a typical day of weekend. While the usage time for the places of Elevator, Front Door, Back Door, and Stairs is short, we ask them to provide the number of usages, instead of the specific time, for these four places, and then we obtain the average staying time by multiplying the number of usages with an approximate time that one person typically spends at one of these four places. The average staying time in minutes for all $11$ places of interest is provided in Table. \ref{average}.

\begin{table}[!ht]
\renewcommand {\arraystretch}{1.0}
\caption{The average usage time in minutes spent at different places collected through the survey \cite{suvery}}
\label{average}
\centering
\begin{tabular}{l c c c}
\Xhline{2\arrayrulewidth}
Places & \textit{Weekday} & \textit{Weeknight} & \textit{Weekend} \\
\Xhline{2\arrayrulewidth}
Dorm Room & $292.8$ & $695.4$ & $820.8$ \\
Halls & $15.77$ & $31.21$ & $16.64$ \\
RA Office & $29.85$ & $22.59$ & $18.83$ \\
Kitchens  & $36.42$ & $69.31$ & $26.18$ \\
Washers/Dryers & $1.08$ & $0.64$ & $4.57$ \\
Vending Machines &  $0.35$ &  $0.39$  & $0.41$ \\
Public Bath & $2.93$ & $1.67$ & $1.43$ \\
Elevator & $0.54$ & $0.29$ &  $0.40$  \\
Front Door & $0.86$ &  $0.46$  & $0.53$ \\
Back Door &  $0.16$ & $0.14$ & $0.21$ \\
Stairs & $1.88$ & $1.13$ &  $1.84$ \\
\Xhline{2\arrayrulewidth}
\end{tabular}
\end{table}

\begin{figure}
\centering
\includegraphics[width=0.5\textwidth,natwidth=610,natheight=642]{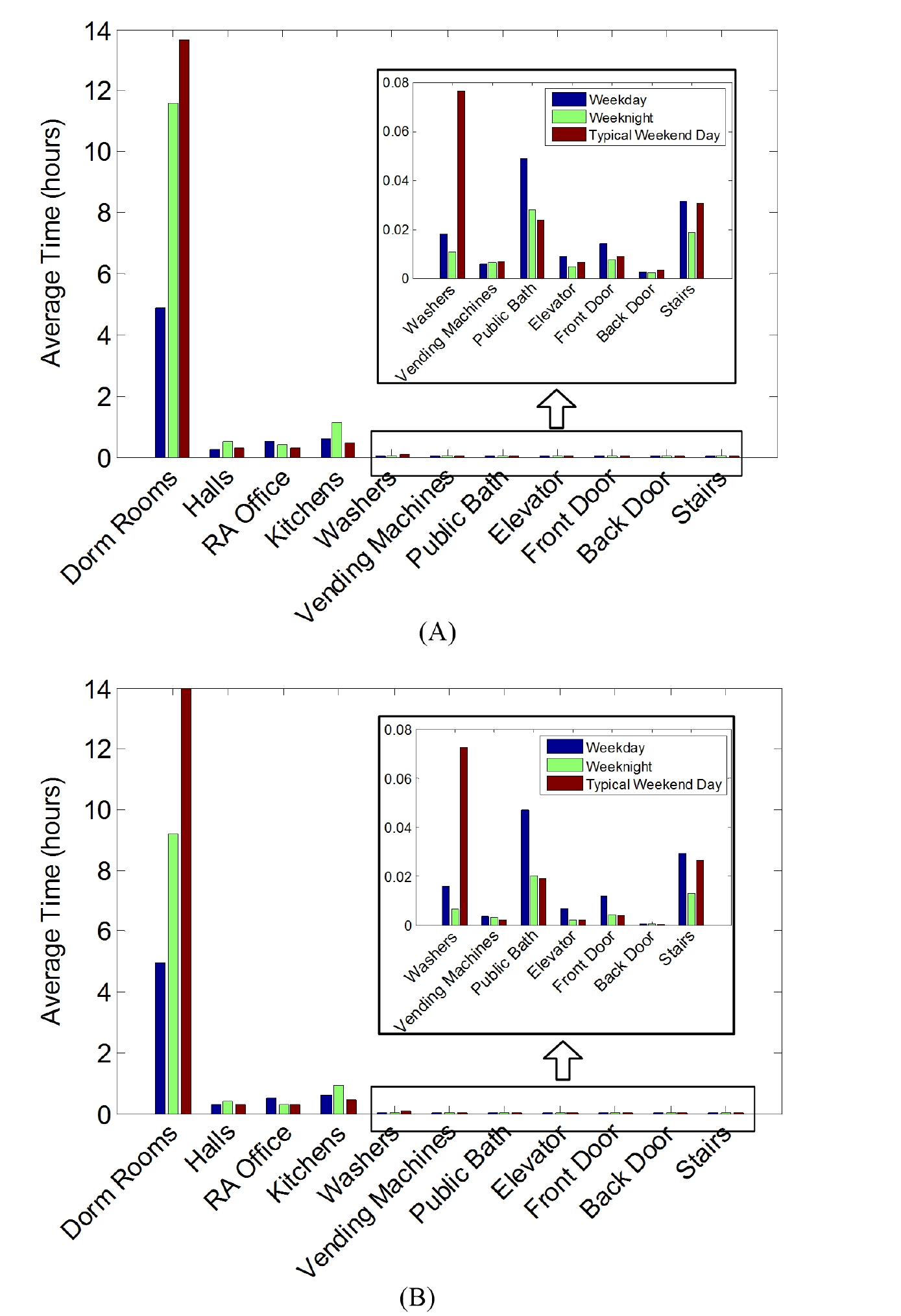}
\caption{Comparison of average usage time of residents at various places between (A) survey data and (B) simulation results in bar graphs.}
\label{fig_average_all}
\end{figure}

\subsubsection{Simulation}
{We use the survey data as the prior knowledge in our probabilistic human mobility model. Specifically, we acquire the prior knowledge $p(s_i | u_j)$ from the survey data, where $s_i$ denotes the $i$-th place ($i=1, 2, \cdots, M$) and $u_j$ denotes the $j$-th type of day ($j=1, 2, \cdots, L$). In this simulation, we have $M = 11$ and $L = 3$. The $u_1$, $u_2$ and $u_3$ represent for a typical day of \textit{Weekday}, \textit{Weeknight} and \textit{Weekend}, respectively. From the survey data, the prior distribution $p(s_i | u_j)$ can be given by:
\begin{align}
p(s_i | u_j) = \frac{\bar{t}_{ij}}{\sum_{k=1}^M \bar{t}_{kj}}
\end{align} 
where $\bar{t}_{ij}$ is the average usage time in Table \ref{average} for the $i$-th place at the $j$-th type of day. We further incorporate this prior knowledge $p(s_i | u_j)$ into our proposed probabilistic human mobility model in Algorithm \ref{algorithm: human dynamic model}. One example of a resident distribution during a typical weekday is shown in Fig. \ref{campus}. For each week, we have $5$ weekdays and weeknights, and $2$ weekend days, and we count the average usage time that each resident spends at $11$ different places for a specific type of day. To reduce the randomness, we simulate the resident distribution for $20$ weeks and average the usage time over these 20 weeks which is shown in Table \ref{simulation}. Compared to the survey data, the simulation results in Table \ref{simulation} show consistent statistical trends to the survey data reported in Table \ref{average}. 

{In Fig. \ref{fig_average_all}, we compare average usage time of residents at various places between survey data and simulation results in bar graphs. Since we use the survey data as prior knowledge, we expect the consistency of statistical trends of average usage time between survey data and simulation results, which can be verified in Fig. \ref{fig_average_all} for each typical data of \textit{Weekday}, \textit{Weeknight}, and \textit{Weekend}. }

\begin{table}[!ht]
\renewcommand {\arraystretch}{1.0}
\caption{The average usage time in minutes spent at different places through simulations}
\label{simulation}
\centering
\begin{tabular}{l c c c}
\Xhline{2\arrayrulewidth}
Places & \textit{Weekday} & \textit{Weeknight} & \textit{Weekend} \\
\Xhline{2\arrayrulewidth}
Dorm Room & $298.42$ & $551.39$ & $836.99$ \\
Halls & $15.94$ & $24.67$ & $16.68$ \\
RA Office & $30.36$ & $17.81$ & $18.85$ \\
Kitchens  & $36.91$ & $54.88$ & $26.59$ \\
Washers/Dryers & $0.95$ & $0.39$ & $4.35$ \\
Vending Machines &  $0.21$ &  $0.19$  & $0.13$ \\
Public Bath & $2.82$ & $1.22$ & $1.14$ \\
Elevator & $0.40$ & $0.12$ &  $0.12$  \\
Front Door & $0.72$ &  $0.26$  & $0.24$ \\
Back Door &  $0.03$ & $0.03$ & $0.02$ \\
Stairs & $1.76$ & $0.78$ &  $1.59$ \\
\Xhline{2\arrayrulewidth}
\end{tabular}
\end{table}
}

\subsubsection{Experimental Data Collection}
\begin{figure}
\centering
\includegraphics[width=0.5\textwidth,natwidth=610,natheight=642]{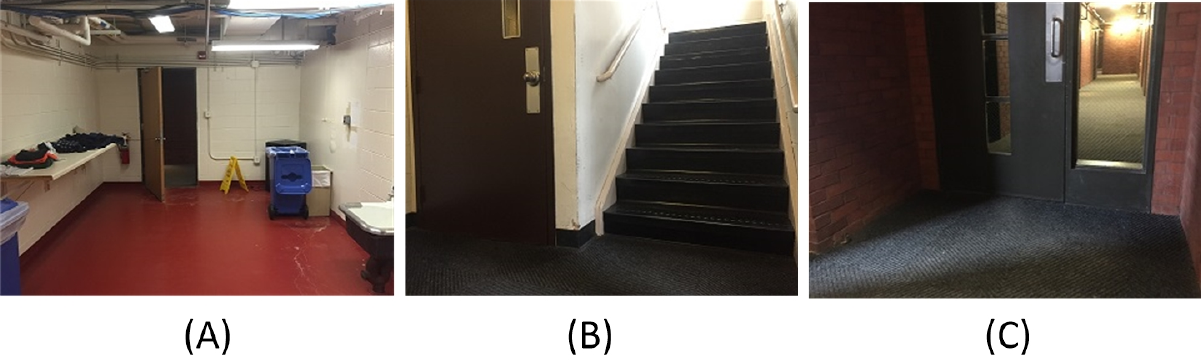}
\caption{{Experimental data collection in (A) laundry room, (B) exit door and (C) back door.}}
\label{real_corners}
\end{figure}

{To further validate the human distributions obtained from the simulation, we also collect real-life data through surveillance cameras that are installed at several specific places \cite{camera}. A limitation of collecting real-life data through surveillance cameras is that it is difficult for us to track individual resident at each specific room due to privacy issues.} Therefore, we only collect camera data at the laundry room and the stair, as shown in Fig. \ref{real_corners}.  Each location is sampled during a weekend, weekday, and weeknight period. To facilitate data collection, a camera (Panasonic model DMC-TZ3) is placed in an unobtrusive location in the area of interest, and left to record low-quality video.  Recording times are typically under two hours, and are generally limited by camera battery life.  Video recording is advantageous over physical observation because many of the stairwells have limited space, and being physically present could cause residents to behave differently. The video recording also allows playback at increased speed, which allows measurements to be made much more efficiently.  A speed of 8x is used, because it is the maximum speed at which people could accurately be counted while passing through.  

\textit{Laundry Room:} The laundry room is expected to have fairly low traffic levels, and a usage time of several minutes for each person. Therefore, for this data, the time when each user entered and left the room was recorded.  A data point was added each time someone entered or left, which allows the occupancy of the room to be known at all points.  From this, the total amount of time, in person-minutes, spent in the room can be calculated by summing all of the time intervals times the occupancy during that interval.  Because the measurements occurred over differing time intervals, they have been adjusted to be comparable over a 1 hour period.  It is important to note that the ``number of uses" statistic counts the number of times an individual entered the laundry room.  Over the course of doing laundry, residents tend to put loads in the washer, leave, and then come back later to move them to the dryer. This means that the same people were counted multiple times, but because this is an inherent part of the room usage, no corrections or modifications were made.

\textit{Stairs:} The average usage time of the stairs in Jonas is very short, so the experimental data collected only recorded the number of usages, not the specific times.  There are three exits to Jonas, two of which are also entrances.  (One door is exit-only.)   Measurements were made at the rear door/rear staircase, and at the front exit-only door/front staircase.  While measurements were not made at the front main entrance, residents who live on any floor except the 3rd (ground level at the front) would likely use the staircase, where they would be recorded. 

\subsubsection{Performance Validation}

\begin{figure}
\centering
\includegraphics[width=0.5\textwidth,natwidth=610,natheight=642]{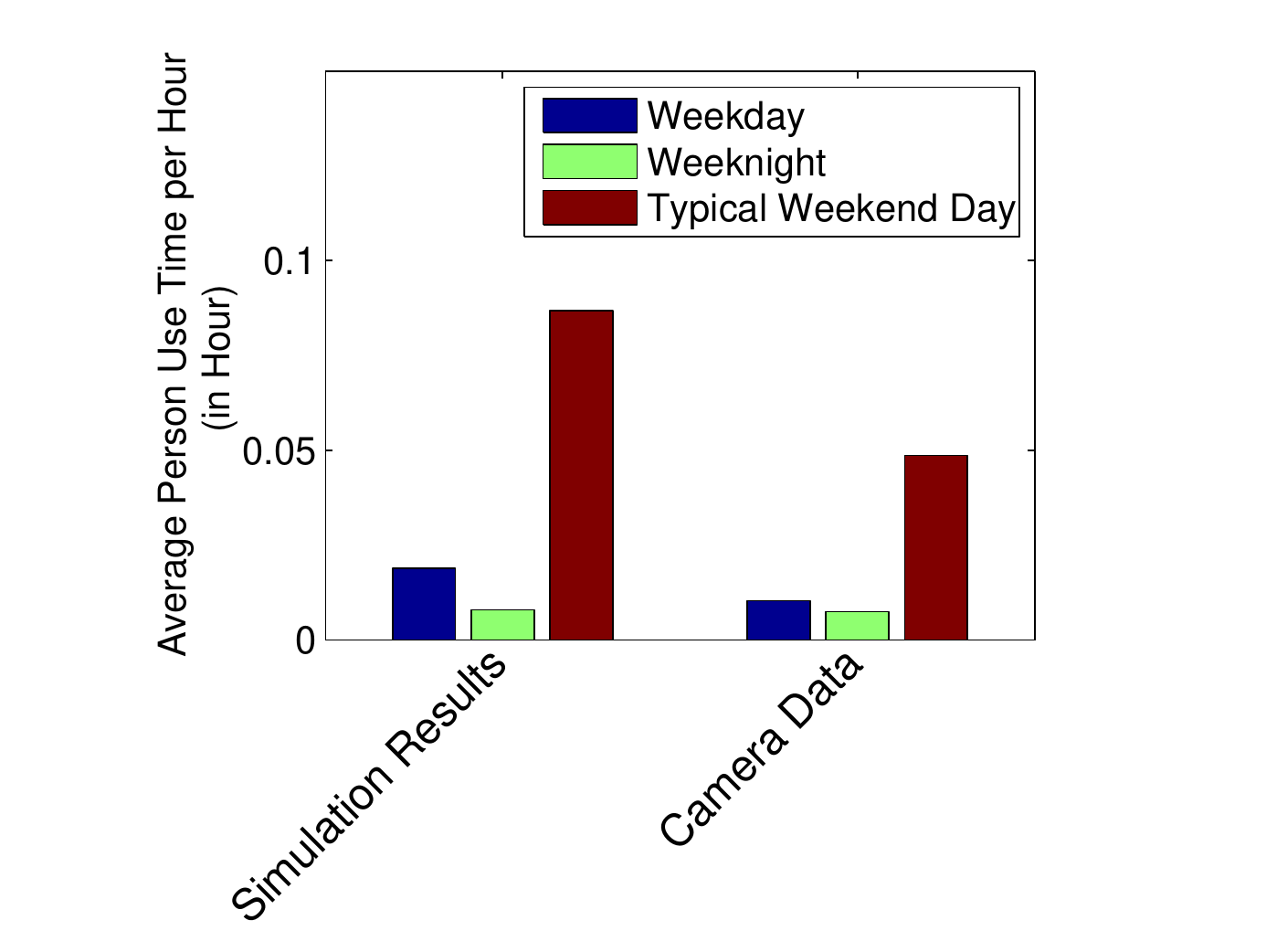}
\caption{Average person usage time per hour at the Laundry Room, comparing simulation results with experimental camera data.}
\label{video_laundry}
\end{figure}

\begin{figure}
\centering
\includegraphics[width=0.5\textwidth,natwidth=610,natheight=642]{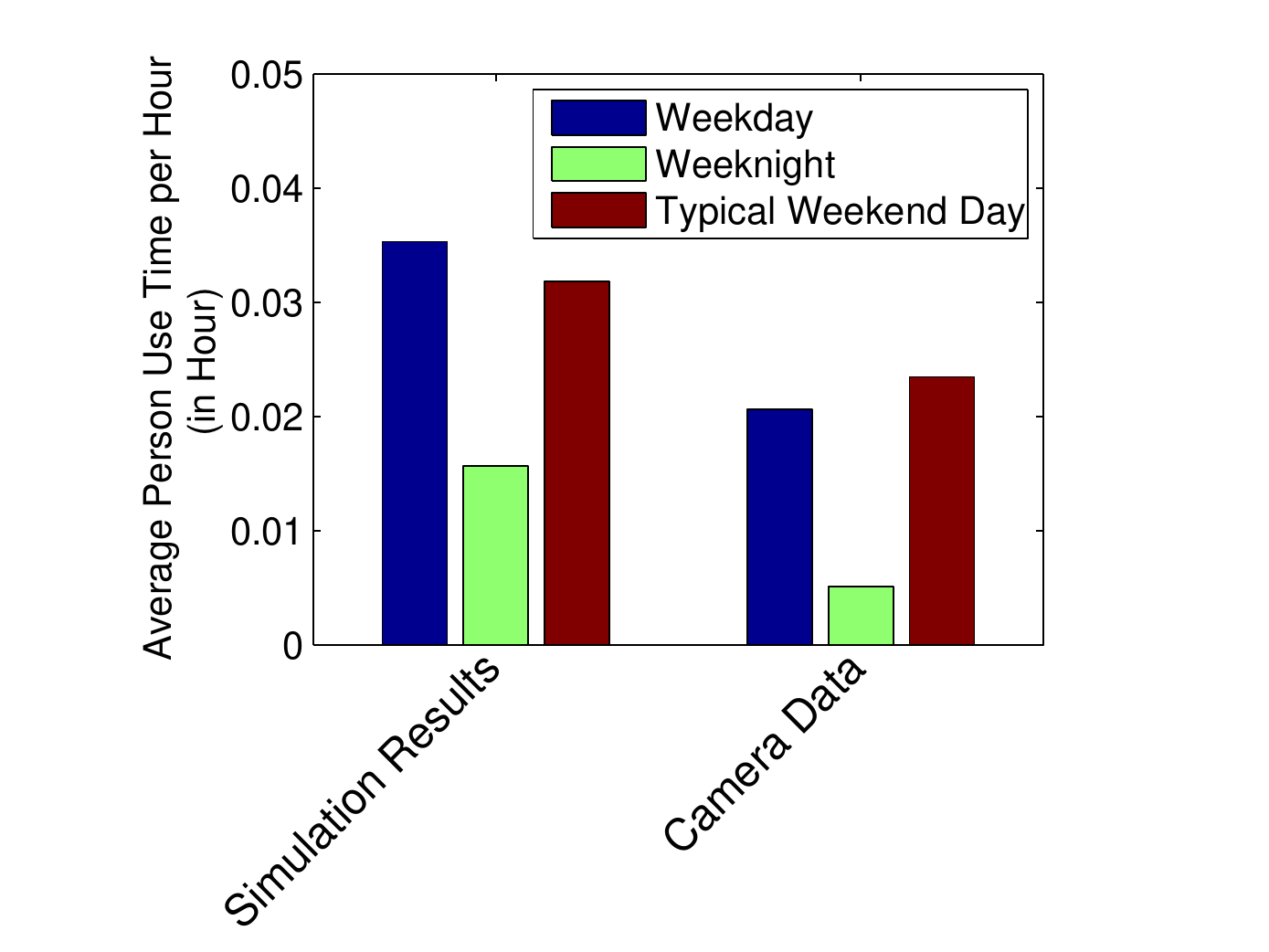}
\caption{Average person usage time per hour at the Stairs, comparing simulation results with experimental camera data.}
\label{video_stair}
\end{figure}

In Fig. \ref{video_laundry}, we show the comparison of average person usage time per hour between simulation results and experimental camera results at Laundry room. Although the experimental data implies that the average person usage time determined experimentally is generally less than the one reported by the survey data and simulation data, the statistical trends of these three methods are consistent. The similar pattern can be also seen in Fig. \ref{video_stair} in which we compare the average person usage time per hour between simulation data and experimental camera data at the Stairs.

The experimental data provides additional insights into the human mobility model on a smaller time scale. There are several reasons why the average usage time based on the experimental data is considerably less than the average usage time obtained from the simulation. Because the experimental data relies on significant extrapolation from the number of hours, there is likely a very high margin of error, and because the front door is split into an exit only and an entrance door, some individuals may have been missed. It is notable that in every case the observed number of uses was lower than that predicted from the simulation and survey data, which implies a systematic rather than random deviation. {Given the number of assumptions made in the calculations, the experimental camera data can validate the simulation results with prior knowledge from survey data. }

\section{Conclusion and Future Work}
In this paper, we investigated human dynamic behavior in indoor environments, which is of great importance but challenging due to the uncertainty of human intention and the complexity of indoor environments. We proposed a new probabilistic human indoor mobility modeling method taking into consideration of both internal and external variables. We validated the proposed human mobility model with two case studies: dynamic human distribution in a shopping mall and average facility usage time in a student dorm building. In the case study of the student dorm building, we incorporated the prior knowledge from the surveys and validated the simulation results using surveillance camera data. Both simulation results and empirical studies have shown the effectiveness of our proposed human indoor mobility model.

Simulations of human dynamic behavior are important and useful for many applications. For example, the current simulations for crowd disaster investigations and emergency evacuation planning usually assume a random evacuee distribution when an emergency happens, which may be unrealistic. Our proposed human indoor mobility model can provide us a more realistic evacuee distribution at emergency and one can better understand the transition of human distribution from normal to emergency situation. In our future work, we plan to employ the proposed human indoor mobility model for robot-assisted emergency evacuation simulations. Meanwhile, we will incorporate the proposed probabilistic human mobility model into the decision making and path planning of mobile service robots. Given the previous records of places visited, one can further estimate the internal variables and predict potential places to visit, therefore improve human-robot interaction and provide better robot-assisted services.

\section*{Acknowledgment}
This work was partially supported by the US National Science Foundation under Grants IIS-1526835 (Tang and He) and IIS-1527016 (Jiang and Guo).

{We would like to thank Stevens' undergraduate students, Alex Carpenter and Robert Whipple, for conducting surveys and collecting experimental data for validation of our proposed human mobility model.}

\bibliographystyle{ieeetr}

\bibliography{ref}

\end{document}

%% file: Macros.tex
\newcounter{appcount}
\newcommand{\appendicesname}
            {Appendix\ \thechapter  \Alph{appcount}}
\newcommand{\bookappendicesname}
            {Appendix\ \Alph{appcount}}
\newcommand{\chapterappendix}[1]
          {\par\setcounter{section}{0}
           \setcounter{equation}{0}
           \setcounter{table}{0}
           \setcounter{figure}{0}
          \addtocounter{appcount}{1}   \renewcommand{\theequation}{\thechapter\Alph{appcount}.\arabic{equation}}
          \renewcommand{\thetable}{\thechapter\Alph{appcount}.\arabic{table}}
          \renewcommand{\thefigure}{\thechapter\Alph{appcount}.\arabic{figure}}
           \setcounter{section}{\arabic{chapter}\Alph{section}}
           \if@openright\cleardoublepage\else\clearpage\fi
           \chapter*{\huge{\appendicesname}\newline\newline \Huge{#1}}
           \addcontentsline{toc}{section}{\thechapter\Alph{appcount} #1}
           \markright{\MakeUppercase{\appendicesname.\ { #1}}}}
\newcommand{\bookappendix}[1]
          {\par\setcounter{section}{0}
           \setcounter{equation}{0}
           \setcounter{table}{0}
           \setcounter{figure}{0}
          \addtocounter{appcount}{1}   \renewcommand{\theequation}{\Alph{appcount}.\arabic{equation}}
           \renewcommand{\thetable}{\Alph{appcount}.\arabic{table}}
             \renewcommand{\thefigure}{\Alph{appcount}.\arabic{figure}}
           \setcounter{section}{\arabic{chapter}\Alph{section}}
           \if@openright\cleardoublepage\else\clearpage\fi
           \chapter*{\huge{\bookappendicesname}\newline\newline \Huge{#1}}
           \addcontentsline{toc}{chapter}{\bookappendicesname #1}
          \markright{\MakeUppercase{\bookappendicesname.\ { #1}}} }
\newcounter{example}
\newcounter{property}
\newcommand{\ben}{\begin{equation}}
\newcommand{\een}{\end{equation}}
\newcommand{\bea}{\begin{eqnarray*}}
\newcommand{\eea}{\end{eqnarray*}}
\newcommand{\bean}{\begin{eqnarray}}
\newcommand{\eean}{\end{eqnarray}}

\newcommand{\bfmu}{\mbox{\boldmath{$\mu$}}}

\newcommand{\bfSig}{{\bf \Sigma}}